\documentclass[prl,superscriptaddress,showpacs,twocolumn]{revtex4-1}
\usepackage{graphicx}
\usepackage{hyperref}
\usepackage{subfigure}
\usepackage{amsmath}
\usepackage{amssymb}

\usepackage{color} 

\newcommand{\ket}[1]{| #1 \rangle}
\newcommand{\bra}[1]{\langle #1 |}

\newcommand{\figref}[1]{Fig.~\ref{#1}}

\begin{document}

\title{Benchmarking a Teleportation Protocol realized in Superconducting Circuits}

\author{M.~Baur}
\altaffiliation{These authors contributed equally to this work.}
\author{A.~Fedorov}
\altaffiliation{These authors contributed equally to this work.}
\author{L.~Steffen}
\author{S.~Filipp}
\affiliation{Department of Physics, ETH Zurich, CH-8093 Zurich, Switzerland}
\author{M.P.~da~Silva}
\affiliation{Raytheon BBN Technologies, Network Centric Systems, 10 Moulton Street, Cambridge, MA 02138 USA}
\affiliation{D\'epartment de Physique, Universit\'e de Sherbrooke, Sherbrooke, Qu\'ebec, J1K 2R1 Canada}
\author{A.~Wallraff}
\affiliation{Department of Physics, ETH Zurich, CH-8093 Zurich, Switzerland}

\date{\today}
\begin{abstract}
Teleportation of a quantum state may be used for distributing entanglement between distant qubits in quantum communication and for quantum computation. Here we demonstrate the implementation of a teleportation protocol, up to the single-shot measurement step, with superconducting qubits coupled to a microwave resonator. Using full quantum state tomography and evaluating an entanglement witness, we show that the protocol generates a genuine tripartite entangled state of all three-qubits. Calculating the projection of the measured density matrix onto the basis states of two qubits allows us to reconstruct the teleported state. Repeating this procedure for a complete set of input states we find an average output state fidelity of 88\%.
\end{abstract}

\pacs{}

\maketitle
Quantum teleportation achieves the transfer of a quantum state from one physical location to another, even if the sender has no knowledge about both the state to be teleported and the location of the receiver~\cite{Bennett1993}. In addition to its use in quantum communication~\cite{Gisin2002}, for example in context of quantum repeaters~\cite{Briegel1998}, quantum teleportation also plays an important role in quantum information processing. For instance, it has been shown that gate teleportation in combination with single qubit operations enables universal and fault-tolerant quantum computation~\cite{Gottesman1999,Zhou2000} in a way closely related to cluster state quantum computation~\cite{Raussendorf2001} which can be understood in a unified conceptual framework~\cite{Childs2005,Aliferis2004,Jorrand2005}. Due to the stringent requirements on the control and read-out fidelity achievable for the multi qubit quantum system, teleportation has so far only been experimentally realized in microscopic degrees of freedom with single photon~\cite{Bouwmeester1997,Boschi1998,Marcikic2003,Jin2010} or continuous variable states~\cite{Furusawa1998,Lee2011} and, more recently, with ions~\cite{Riebe2004,Barrett2004,Riebe2007,Olmschenk2009}. Early experiments have also been performed with spins by use of nuclear magnetic resonance technique~\cite{Nielsen1998}.

Rapid progress towards increasing the fidelity of control and read-out operations has also been made in macroscopic systems, such as superconducting circuits. Recently, single qubit operations have been carried out with fidelities of up to $99\%$~\cite{Chow2010} and two-qubit entangled states have been generated with fidelities of up to $95\%$~\cite{DiCarlo2009, Chow2011}. Two-qubit algorithms, such as the Deutsch Jozsa and Grover search algorithm, have been demonstrated~\cite{DiCarlo2009,Yamamoto2010} and tripartite entangled states (GHZ and W) have been realized~\cite{DiCarlo2010,Neeley2010a}.

\begin{figure}[!b]
  \includegraphics[width=0.9\columnwidth]{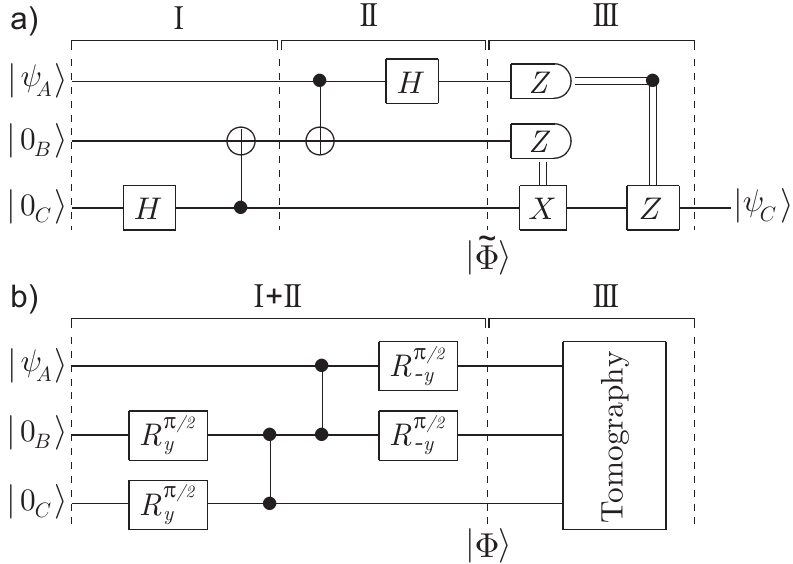}
  \caption{a) Circuit diagram of the standard protocol to teleport the state $\ket{\psi_A}$ of qubit A to qubit C. Here, $H$ is the Hadamard gate, $Z$ and $X$ are the Pauli matrices $\sigma_z$ and $\sigma_x$. The CNOT gate is represented by a vertical line between the control qubit ($\bullet$) and the target qubit ($\oplus$). b) The circuit implemented in this experiment with controlled phase gates, indicated by vertical lines between the relevant qubits ($\bullet$), and single qubit rotations $R_{\hat{n}}^\theta$ of angle $\theta$ about the axis $\hat{n}$.}
  \label{fig:circuit}
\end{figure}
Here, we demonstrate a scheme to fully characterize the operation of a teleportation circuit, without making use of single shot read-out~\cite{Mallet2009} or real-time feed-back~\cite{Doherty1999}, both of which are challenging to realize simultaneously in a three-qubit superconducting quantum processor at the current state of the art. Even without explicitly realizing these steps, which will be essential in future applications, our benchmarking process is able to provide crucial information on the operation and fidelity of the teleportation process up to the measurement step and thus presents an important achievement towards making use of teleportation in quantum processors realized in superconducting circuits.

In the standard protocol, teleportation is performed using non-local quantum correlations combined with classical communication. In this scheme (see \figref{fig:circuit}a)) the sender is in possession of qubit A in an arbitrary state $\ket{\psi_A}$. In the first step (I), a maximally entangled pair is generated, e.g.~using a Hadamard (H) gate followed by a controlled-not (CNOT) gate, and shared between the sender (qubit B) and the receiver (qubit C). In the second step (II) the sender applies a CNOT gate on his two qubits followed by a H gate on qubit A generating an entangled three-qubit state $\ket{\tilde\Phi}$. In step III, the sender performs a measurement on his two qubits, which combined with step II is equivalent to a measurement performed in the Bell basis. He then sends the digital results to the receiver over a classical communication channel. Depending on these results, the receiver applies one of four unitary operations to his qubit to transform the state $\ket{\psi_C}$ of qubit C into the state $\ket{\psi_A}$, completing the teleportation.

In our approach using superconducting qubits we realize steps I and II by combining single qubit rotations and two-qubit controlled phase gates, as illustrated in \figref{fig:circuit}b), to create the entangled state
\begin{align}
    \label{eq:teleportationstate}
    \ket{\Phi}&=\dfrac{1}{2}\left\{\ket{0_A0_B}
    \otimes\ket{\psi_C}+\ket{0_A1_B}\otimes(-\sigma_x)\ket{\psi_C}\right.\\
    &\quad\left.{}+\ket{1_A0_B}\otimes(-\sigma_z)
    \ket{\psi_C}+\ket{1_A1_B}\otimes(-i\sigma_y)\ket{\psi_C}\right\}.\nonumber
\end{align}
In this notation, it becomes clear that a measurement of qubits A and B collapses qubit C onto one of four possible states. If the measurement outcome is 00, 01, 10, or 11, qubit C is projected to the states $\ket{\psi_C}$, $-\sigma_x\ket{\psi_C}$, $-\sigma_z\ket{\psi_C}$ or $-i\sigma_y\ket{\psi_C}$, respectively. Instead of performing single qubit measurements on qubits A and B in step III, we analyze the three-qubit entangled state $\ket{\Phi}$ with full quantum state tomography and reconstruct the teleported state by calculating the projection of qubits A and B onto the basis states $\ket{0_A0_B}$, $\ket{0_A1_B}$, $\ket{1_A0_B}$ and $\ket{1_A1_B}$. We then characterize the transfer of the input state $\ket{\psi_A}$ to qubit C by performing process tomography conditioned on the projection onto the basis states of qubits A and B.

\begin{figure}[!b]
  \includegraphics[width=0.95\columnwidth]{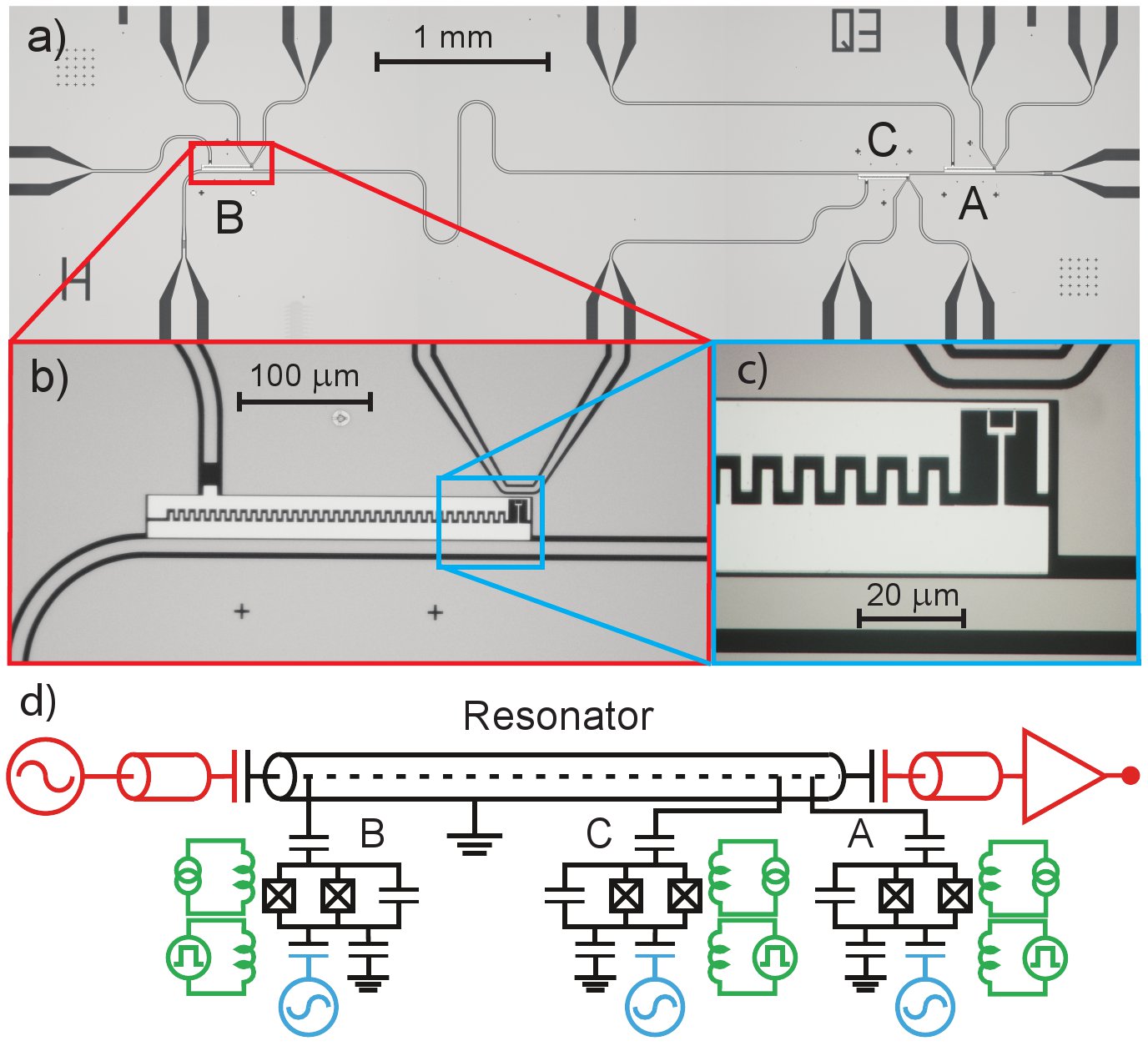}
  \caption{a) Optical microscope image of the sample with three qubits coupled to a coplanar wavequide resonator with individual local microwave and magnetic flux-bias lines for each qubit. b) and c) show a close up of qubit B. d) Lumped element circuit diagram of the sample and measurement setup.}
  \label{fig:sample}
\end{figure}

An optical microscope image of our sample consisting of three transmon qubits (A,B,C)~\cite{Koch2007} dispersively coupled to a microwave transmission line resonator is shown in \figref{fig:sample} combined with the corresponding lumped element circuit diagram.
The resonator has a bare resonance frequency of $\nu_{\mathrm{r}}=8.625\,\mathrm{GHz}$ and a quality factor of $3300$. It acts as a coupling bus~\cite{Majer2007} between the qubits and allows to perform joint three-qubit readout by measuring its transmission~\cite{Filipp2009b}. The qubits have a slightly anharmonic ladder-type energy level structure. The first two levels are used as the computational qubit states $\ket{0}$ and $\ket{1}$, while the second excited state $\ket{2}$ is used to perform two-qubit operations. For optimal coherence, we designed qubits with maximal transition frequencies smaller than $\nu_\mathrm{r}$ and anharmonicities big enough to address the first excited state without exciting higher states. From spectroscopy we extract the maximum transition frequencies $\nu^\mathrm{max}_\mathrm{A,B,C}=\{6.714,\,6.050,\,4.999\}\,\mathrm{GHz}$, charging energies $E_c/h=\{0.264,\,0.296,\,0.307\}\,\mathrm{GHz}$ and coupling strength to the resonator $g/2\pi=\{0.36,\,0.30,\,0.34\}\,\mathrm{GHz}$. To maximize coherence, we independently tune each qubit transition frequency to $\nu^\mathrm{max}$ using superconducting coils mounted underneath the chip. At this optimal bias point, we find energy relaxation times $T_1=\{0.55,\,0.70,\,1.10\}\,\mathrm{\mu s}$ and phase coherence times $T_2^*=\{0.45,\,0.60,0.65\}\,\mathrm{\mu s}$.

With two local control lines at each qubit, shown in \figref{fig:sample}b), we perform arbitrary single qubit operations with fidelity greater than 98\% \cite{Chow2010}. Resonant microwave pulses applied to the open-ended transmission line realize single-qubit rotations about the x and y axis~\cite{Gambetta2011a}. Nanosecond time-scale current pulses applied to the transmission line passing by the SQUID loop control the qubit transition frequency realizing z-rotations.

The controlled phase (C-Phase) gate is implemented using the avoided level crossing between $\ket{11}$ and $\ket{20}$~\cite{Strauch2003}. A fast magnetic flux pulse first shifts $\ket{11}$ non-adiabatically into resonance with $\ket{20}$. The system then oscillates between the two states with twice the frequency of the cavity mediated transverse coupling strength of $J_{11,20}^{AB}=36\,\mathrm{MHz}$ (between qubits A and B) and $J_{11,20}^{BC}=23\,\mathrm{MHz}$ (between qubits B and C). After an interaction time $t=2\pi/2 J_{11,20}$, the system returns to the initial state $\ket{11}$ with an additional phase factor -1. No conditional phase is picked up by the other basis states $\ket{00}$, $\ket{01}$ or $\ket{10}$. Dynamic single qubit phases are canceled by adjusting the rotation axes of all subsequent single qubit operations appropriately. The full two-qubit gates between A and B, and between B and C are completed in $14\,\mathrm{ns}$ and $22\,\mathrm{ns}$, respectively.
\begin{figure*}[!ht]
  \centering
  \includegraphics[width=0.85\textwidth]{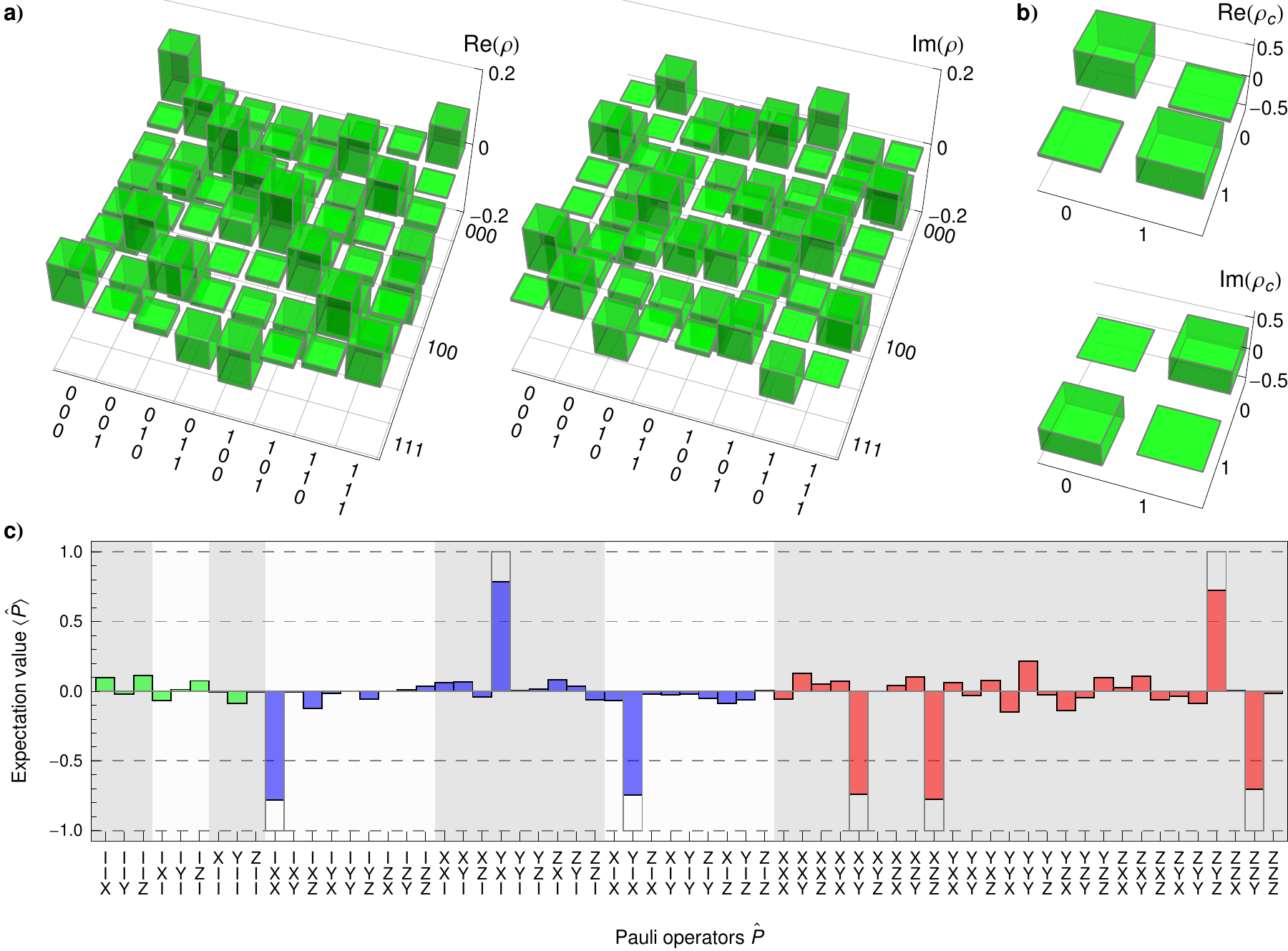}
  \caption{a) Real and imaginary part of the measured three-qubit density matrix $\rho_\mathrm{m}$ when applying the circuit shown in Fig.~\ref{fig:circuit}b) to the input state $\ket{\psi_A}=(\ket{0_A}-i\ket{1_A})/\sqrt{2}$. b) Teleported single qubit state at qubit C found by projecting $\rho_\mathrm{m}$ from a) onto $\ket{0_A0_B}$ and tracing out qubits A and B. c) Pauli sets for the state shown in a).}
  \label{fig:state_tomograms}
\end{figure*}

We characterize the teleportation process realized according to the circuit diagram in \figref{fig:circuit}b) with full quantum state tomography using joint read-out~\cite{Filipp2009b} for a complete set of input basis states $\ket{\psi_A}=\{\ket{0_A}$, $\ket{1_A}$, $\ket{-_A}$, $\ket{+_A}\}$. Here, $\ket{-_A}=(\ket{0_A}-i\ket{1_A})/\sqrt{2}$ and $\ket{+_A}=(\ket{0_A}+\ket{1_A})/\sqrt{2}$. In \figref{fig:state_tomograms}a we show the measured density matrix $\rho_\mathrm{m} = \ket{\Phi_\mathrm{m}}\bra{\Phi_\mathrm{m}}$ of the three qubit state $\ket{\Phi_\mathrm{m}}$ generated for the input state $\ket{-_A}$ as an example. We apply a maximum likelihood method~\cite{Smolin2011} to ensure that $\rho_\mathrm{m}$ is physical and determine the fidelity $F=\bra{\Phi}\rho_\mathrm{m}\ket{\Phi}=0.78$ with respect to ideal state $\ket{\Phi}$. We note that for this particular input state, $\ket{\Phi}$ is a cluster state useful for one way quantum computation~\cite{Raussendorf2001}. For the input states $\ket{0_A}$, $\ket{1_A}$ and $\ket{+_A}$ the fidelities are $82\%$, $79\%$ and $80\%$, respectively, comparable to the best fidelities of three-qubit entangled states realized in superconducting qubits so far~\cite{DiCarlo2010,Neeley2010a}. Also, the measured correlations (colored bars) present in $\rho_{\mathrm{m}}$ expressed in terms of Pauli sets, displaying the expectation values of all nontrivial tensor products $P$ of identity $I$ and Pauli operators $X, Y, Z$ for three-qubits, are in good agreement with the expected ones (wireframe), see \figref{fig:state_tomograms}c).

Generally, the ideal three-qubit state $\ket{\Phi}$ generated by the circuit is tripartite entangled as can be verified by calculating the three tangle (residual tangle) defined for pure states~\cite{Coffman2000}. Only for $\ket{0_A}$ and $\ket{1_A}$ the output state $\ket{\Phi}$ remains biseparable. To quantify the amount of entanglement in the measured state $\rho_\mathrm{m}$, we estimate the three tangle $\tau_3(\rho)$ for mixed states via the convex-roof extension~\cite{Uhlmann1998}. The values $\tau_3(\rho)=\{0.49,\,0.52\}>0$ demonstrate that GHZ-type tripartite entanglement was prepared for $\ket{\psi_A}=\{\ket{-},\,\ket{+}\}$. If we only want to verify that $\rho_\mathrm{m}$ contains tripartite entanglement without distinguishing between the GHZ and W class, we can use a witness operator $\mathcal{W}=\alpha \mathbb{I}-\ket{\Phi}\bra{\Phi}$~\cite{Bourennane2004}. Here, $\alpha$ is the maximal squared overlap of any biseparable state with $\ket{\Phi}$, which yields $0.5$ for $\ket{\pm_A}$. For all biseparable states we find $\mathrm{Tr}(\mathcal{W}\rho) \geq 0$, whereas for the ideal tripartite entangled state $\rho=\ket{\Phi}\bra{\Phi}$ we find $\mathrm{Tr}(\mathcal{W}\rho) = \alpha-1$. According to this criterion $\mathrm{Tr}(\mathcal{W}\rho_{\rm{m}}) = -0.28 < 0$ the measured state shown in \figref{fig:state_tomograms}a) clearly has tripartite entanglement. As derived in~\cite{Eisert2007}, the expectation value of the witness operator also directly leads to a lower bound to the generalized robustness of entanglement. It measures the minimal amount of mixing of $\rho_\mathrm{m}$ with an arbitrary density matrix $\sigma$ such that $\rho_\mathrm{m}+s\sigma$ is separable, for which we find $s\geq0.56$.
\begin{figure}[!t]
  \centering
  \includegraphics[width=0.87\columnwidth]{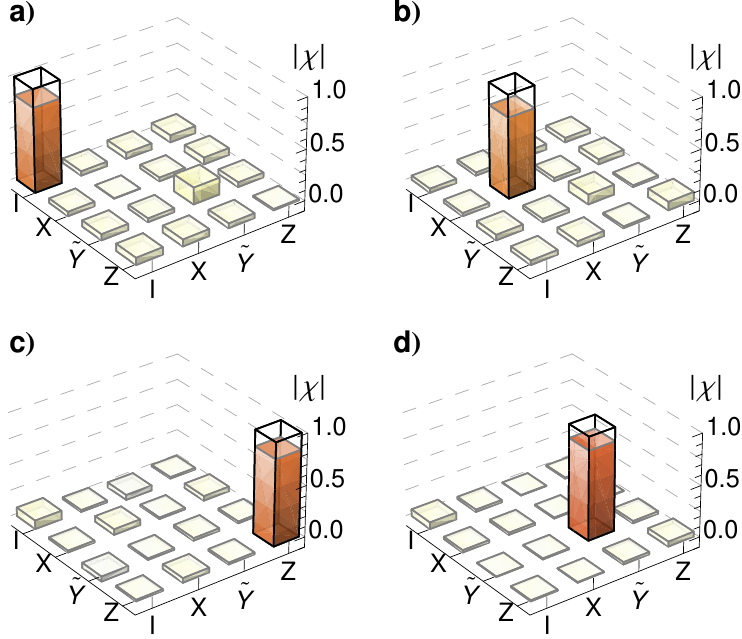}
  \caption{Absolute value of the $\chi$-matrix representation for the teleportation process which transfers the state of qubit A to qubit C for qubit A and B being projected to a) $\ket{0_A0_B}$, b) $\ket{0_A1_B}$, c) $\ket{1_A0_B}$ and d) $\ket{1_A1_B}$. $I$ is the identity matrix, $X$, $\tilde{Y}$ and $Z$ the Pauli matrices $\sigma_x$, $-i\sigma_y$ and $\sigma_z$ respectively.}
  \label{fig:process_matrices}
\end{figure}

To determine the fidelity of the teleportation process up to the measurement, we calculate the projection of $\rho_\mathrm{m}$ onto the four basis states $\ket{0_A0_B}$, $\ket{0_A1_B}$, $\ket{1_A0_B}$, and $\ket{1_A1_B}$. The state of qubit C is then reconstructed by tracing out qubits A and B and renormalizing the density matrix to $\rho_C =\mathrm{Tr}_{AB}(P_{ij}\rho_\mathrm{m}P_{ij}^\dagger)/\mathrm{Tr}(P_{ij}\rho_\mathrm{m})$,
where $P_{ij}$ are the projectors $\ket{i_Aj_B}\bra{i_Aj_B}\otimes\mathbb{I_C}$. When projecting onto $\ket{0_A0_B}$, this state is expected to be identical to the input state $\ket{\psi_A}$. Fig.~\ref{fig:state_tomograms}b) shows $\rho_C$ reconstructed from the measured data for the input state $\ket{-_A}$ with a fidelity of 88\%. For the other three projections, we find the resulting states of qubit C $-\sigma_x\ket{-_C}$, $-\sigma_z\ket{-_C}$ and $-i\sigma_y\ket{-_C}$ with respective fidelities of $82\%$, $82\%$ and $89\%$.

To fully characterize the teleportation circuit, we have performed quantum process tomography of the state transfer by repeating the procedure described above for $\ket{0_A}$, $\ket{1_A}$, $\ket{-_A}$, and $\ket{+_A}$. With the known input states and the reconstructed state of qubit C after teleportation, we calculate the process matrix $\chi$ of the transfer. The extracted matrices $\chi_\mathrm{m}$ clearly demonstrate that the effective processes acting on the target qubit during teleportation are the unitary operations expected from Eq.~\eqref{eq:teleportationstate}, see \figref{fig:process_matrices}. Since the $\chi_\mathrm{m}$ have only small imaginary elements $< 0.07$, we display the absolute value of $\chi_\mathrm{m}$ for the different calculated projections on qubits A and B to emphasize the deviations from the ideal matrices $\chi_\mathrm{t}$ indicated by wireframes.  The corresponding process fidelities $F_\mathrm{p}=\mathrm{Tr}(\chi_\mathrm{m}\cdot\chi_\mathrm{t})$ are 82\%, 78\%, 84\%, 87\%, yielding $83\%$ averaged over all measurement outcomes. The average output state fidelity $\bar F=(2F_\mathrm{p}+1)/3$ is $88\%$, $85\%$, $89\%$, $91\%$ for each individual process, and $88\%$ on average.

In summary, we have benchmarked a teleportation algorithm by tomographic reconstruction of the three-qubit entangled state generated by the circuit up to the single qubit measurements. Using an entanglement witness, we showed that this state has genuine tripartite entanglement. We determined the fidelity of the teleported state by reducing the density matrix with projection and find a high average output state fidelity suggesting that full teleportation above the classical limit of $\bar F=2/3$ is likely to become possible in the near future with superconducting qubits by combining our setup with a high fidelity single shot readout, e.g.~with Josephson bifurcation~\cite{Mallet2009} or parametric amplifiers~\cite{Vijay2011}, and feedback~\cite{Doherty1999}.

\begin{acknowledgments}
We acknowledge fruitful discussions with A. Blais. This work was supported by the Swiss National Science Foundation (SNF),
the EU IP SOLID, and ETH Zurich.
\end{acknowledgments}

\vspace{-5mm}


\end{document}